# Acceptance & Rejection

**Khashayar Irani**
**Birkbeck College, University of London**
**k.irani@mathematicallogic.com**
**September 2026**

**Abstract**

In this paper, we argue that acceptance and rejection are classical speech acts that express two opposite polar positions toward propositional content and can be systematically represented within a classically outlined bilateral formal model.[1] Acceptance occupies the positive pole, represented by +A, whereas rejection occupies the negative pole, represented by −A. Their classical relationship is mediated by negation because acceptance of A corresponds to rejection of ¬A, while rejection of A corresponds to acceptance of ¬A, represented by +A ↔ −(¬A) and −A ↔ +(¬A).[2] Coherency and incoherency are introduced as supplementary notions to clarify this relationship and its connection with reflexivity. A further result of the manuscript is the methodical distinction between rejection and negation namely, rejection expresses negative polarity toward A, whereas ¬ operates upon propositional content.[3]

Acceptance and rejection are classical speech acts expressing two opposite positions that may be taken toward propositional content namely, acceptance occupies the positive pole, whereas rejection occupies the negative pole.[4] If A represents the proposition "The Earth is round",

[1] This concise investigation is the first in a series of papers contributing to a unified project whose ultimate aim is to demonstrate, through a particular formulation of bilateralism, that classical logic, like other well-behaved logical systems, can support a well-mannered proof system in which classical operators, particularly the traditionally problematic case of negation, can operate within a classical environment without generating the proof-theoretic difficulties commonly associated with them.

[2] The mediating role of negation, in the sense developed in this analysis, constitutes one among a number of considerations that lead us to strongly hold that negation plays a structural role in logic, particularly in classical logic. By systematically mediating between the positive and negative polar positions associated with acceptance and rejection, negation appears to perform a function extending beyond that of an ordinary logical constant. Nevertheless, the structural status of negation constitutes a complex research area in its own right. We shall therefore leave the methodical formulation, formal development, and detailed argumentation for this thesis to a separate and independent investigation.

[3] In our view, the distinction between rejection and negation is consequently essential to understanding the classical relation between acceptance and rejection.

[4] We believe that in a fully formulated bilateral setting, through the employment of speech-act elements such as acceptance and rejection, one can provide a classical interpretation of logical inferentialism. For an introduction to logical inferentialism and the challenges that classical logic faces within a logical inferentialist framework,

acceptance of A may be represented by +A, while rejection of A may be represented by −A. In both expressions the propositional content remains A; what differs is the position occupied toward it. The signs + and − therefore provide a formal abstraction of the polarity expressed by these two speech acts. They should not themselves be identified with actual acts performed by speakers. The notation does not specify who accepts or rejects A, when the act occurs, or the circumstances in which it is performed. Instead, + and − abstract the positive and negative positions expressed by acceptance and rejection and represent those positions within a bilateral formal setting.[5] This distinction allows the bilateral model to preserve the structural opposition between the two classical speech acts while separating their formal representation from the contingent features of ordinary linguistic behaviour. Most importantly, it establishes that propositional content and polarity are distinct. A remains the same propositional content in both +A and −A; only its polar position changes. The primary task is therefore to explain how these two opposite speech acts can provide the basis for two formally opposed positions without reducing either position to the other.[6]

However, prior to proceeding any further, a noteworthy point must be stated at the outset of this work, namely, that within this project our employment of acceptance and rejection does not constitute a system of illocutionary logic. A classical example of such a system is John Rogers Searle's and Daniel Vanderveken's (1985) book *Foundations of Illocutionary Logic*, in which their main development is to study the logical relations among speech acts. They clarify their primary aim as:

---

see Khashayar Irani (2025). Equally, for a general introduction to the notion of speech acts, see Mitchell Green's (2021) article on this subject and the references provided therein.

[5] Although our formalisation of bilateralism in this manuscript is to some extent similar to that proposed by Ian Rumfitt (2000), the two approaches differ in a number of important respects. One significant distinction concerns the formation of compound formulae within their respective bilateral formal languages. In Rumfitt's formalism, the sign governing a compound formula is placed externally, while its component formulae remain unsigned; thus, an accepted conjunction is represented as +(A ∧ B). By contrast, as will be shown in the subsequent paragraphs, where the rationale for this feature will also be elucidated, our bilateral language preserves both the external sign of the completed compound and the original internal signs of its constituents, yielding structures such as +(+A ∧ +B), −(−A ∧ −B), +(+A ∧ −B), or −(−A ∧ +B). There are, of course, further distinctions between our bilateral formalism and Rumfitt's, both in their formal features and their broader philosophical significance. A comprehensive comparison, however, lies beyond the scope and aims of the present enquiry and is therefore left for another occasion.

[6] The significance of acceptance and rejection extends beyond their treatment as speech acts, since these notions have also played an important role in proof-theoretic investigations of inference and logical consequence. A representative example is Tor Sandqvist's (2012) paper, *Acceptance, Inference, and the Multiple-Conclusion Sequent,* which develops an interpretation of multiconclusion sequents in terms of rational acceptance rather than truth and falsity. Sandqvist interprets multiconclusion sequents as a form of metainference, whereas singleconclusion sequents represent inferences from sentences to sentences, multiconclusion sequents represent a certain kind of inference between singleconclusion sequents. The resulting semantics is intended to render the standard structural rules of reflexivity, monotonicity, and transitivity both sound and complete. However, particularly relevant to the present enquiry is Sandqvist's observation that earlier attempts to account for multiconclusion sequents without appealing to truth and falsity had typically employed primitive notions of both acceptance and rejection, whereas his own treatment seeks to proceed through acceptance alone. For technical reasons, his account is restricted to sequents with non-empty succedents.

> A theory of illocutionary logic of the sort we are describing is essentially a theory of illocutionary commitment as determined by illocutionary force. The single most important question it must answer is this: Given that a speaker in a certain context of utterance performs a successful illocutionary act of a certain form, what other illocutions does the performance of that act commit him to?[7]

Furthermore, the bilateral formal treatment and philosophical employment of acceptance and rejection advanced here should also be distinguished from Anna Gomolińska's (1998) paper *On the Logic of Acceptance and Rejection*, in which acceptance and rejection are developed within a nonmonotonic formalism concerned with representing the states of knowledge of an introspective agent and the decisions made about available information, as she introduces the project in the following words:[8]

> The logic of acceptance and rejection (AEL2) is a nonmonotonic formalism to represent states of knowledge of an introspective agent making decisions about available information. Though having much in common, AEL2 differs from Moore's autoepistemic logic (AEL) by the fact that the agent not only can accept or reject a given fact, but he/she also has the possibility not to make any decision in case he/she does not have enough knowledge.[9]

The emphasis upon illocutionary commitment in Searle's and Vanderveken's account, together with the epistemic and nonmonotonic orientation of Gomolińska's AEL2, marks important differences from the bilateral theory advanced here.[10] The formal concern of the present theory is neither to determine which further illocutionary acts a speaker is committed to by performing a particular speech act nor to represent the states of knowledge and decisions of an introspective agent. Rather, it abstracts from the acts of acceptance and rejection two formally opposed polar positions toward propositional content. Hence, the theory investigates the structural relations between these positive and negative positions independently of the particular speaker, context of utterance, further illocutionary commitments, or epistemic states associated with their performance. Acceptance and rejection therefore enter the bilateral model as classical speech acts whose opposing polarities provide the formal basis for the positive and negative positions represented by + and −.

---

[7] Searle & Vanderveken (1985) Page: 6

[8] The principal similarity with Gomolińska's account lies in treating acceptance and rejection as formally distinguishable rather than reducing rejection to ordinary propositional negation. Gomolińska's AEL2, nevertheless, develops this distinction within a specifically autoepistemic and nonmonotonic framework. Its purpose is to represent states of knowledge of an introspective agent who may accept a fact, reject it, or refrain from either decision when sufficient knowledge is unavailable. Accordingly, AEL2 employs separate operators for acceptance and rejection and organises epistemic states around sets of accepted and rejected formulae. This differs substantially from the bilateral approach developed here, since Gomolińska's primary concern is the representation of an agent's epistemic state and decision-making rather than the abstraction of opposing polar positions from classical speech acts. Nonetheless, both approaches recognise that acceptance and rejection require formally distinct representation, making Gomolińska's work an important related treatment while differing significantly in its formal purpose, philosophical interpretation, and epistemic orientation.

[9] Gomolińska (1998) Page: 233

[10] For a brief but detailed introduction to epistemic logic, see Rasmus Rendsvig's, John Symons', & Yanjing Wang's (2025) article and the references they provide on this subject.

To further clarify the formal opposition between acceptance and rejection, the notions of coherency and incoherency are introduced as supplementary concepts.[11] Within this account, coherency concerns a properly structured relation between acceptance and rejection in which their opposition is preserved rather than collapsed. A configuration is coherent when the inferential movement between the positive and negative positions is mediated by negation i.e. acceptance of A is coherently related to the inferential rejection of ¬A, while rejection of A is coherently related to the inferential acceptance of ¬A.[12] Incoherency, by contrast, arises when acceptance and rejection are imposed directly upon the same propositional content, so that A occupies contrary polar positions without negation mediating their opposition. Coherency and incoherency are therefore not independent subjects alongside acceptance and rejection. They are supplementary notions used to explain more precisely how the two classical speech acts can be formally organised. Consider A = “The building is open”. Acceptance of this proposition is represented by +A, while rejection of precisely the same proposition is represented by −A. A direct inference such as $+A \vdash -A$ moves from acceptance to rejection while leaving A unchanged; similarly, $-A \vdash +A$ moves from rejection to acceptance of identical propositional content. Such inferences are incoherent because the positive and negative poles are made directly interchangeable with respect to the same A. Coherency, by contrast, preserves their opposition by requiring a systematic mediation between them rather than permitting either pole simply to collapse into its contrary.

The structural notion of reflexivity helps clarify why this difference matters.[13] Reflexivity requires the relevant propositional content to retain a stable identity rather than being treated as reflexively interchangeable with a contrary polar position toward itself. The direct inferences $+A \vdash -A$ and $-A \vdash +A$ undermine this stability because A remains unchanged while the position toward A is converted directly into its opposite. If A represents “The train has arrived”, $+A \vdash -A$ would inferentially move directly from acceptance of “The train has arrived” to rejection of that identical proposition. Nothing has changed in the propositional content to mediate the change from one pole to the other. In this sense, incoherency undermines reflexivity because it allows acceptance and rejection to be assigned directly to the same A as though the distinction between positive and negative positions could be disregarded. Coherency upholds reflexivity because it preserves the distinction between the two positions while relating them through a corresponding change in propositional content supplied by negation. Reflexivity, like coherency and incoherency, therefore in this enquiry serves an explanatory role in the analysis of acceptance and rejection. Acceptance and rejection remain primary; coherency describes their inferentially structured relationship, incoherency identifies the inferential failure of that structure, and reflexivity helps explain why the distinction between these cases is formally significant.

[11] As fundamental epistemic concepts possessing a solid logical foundation, the notions of coherency and incoherency developed in this study can be extended to a range of philosophical domains beyond the present enquiry, beginning with the analysis of logical consequence and the theory of meaning. However, our present characterisation of coherency and incoherency should not be identified with the coherence theory of truth. According to James O. Young (2026), “A coherence theory of truth states that any proposition is true if and only if it coheres with some specified set of propositions.” Our use of coherency and incoherency concerns instead the formal organisation of the opposing positions associated with acceptance and rejection.

[12] From a warrant-theoretic and bilateral point of view, the notion of coherency developed in the present enquiry can give rise to a particular form of inferential warrant or entitlement, which, for future reference, we shall call warrant-coherency. Warrant-coherency concerns the basic inferential entitlement of the relation between the positive and negative positions associated with acceptance and rejection. Accordingly, acceptance of A can coherently provide the inferential warrant to infer the rejection of ¬A, while rejection of A coherently provides the inferential warrant to infer the acceptance of ¬A.

[13] For further discussion of the centrality of reflexivity and the accompanying notions of monotonicity and cut in classical logic, see Rumfitt (2015).

Negation supplies the mediation required for the coherent relationship between acceptance and rejection, but its formal role must be distinguished from the negative position expressed by rejection.[14] If A is "The building is open", the formal negation $\neg A$ will be rendered throughout this manuscript as "It is not the case that the building is open". This wording is deliberately preferred to "The building is not open". The latter is perfectly intelligible ordinary English, but the expression "is not" does not clearly display the formal operation performed by $\neg$. In $\neg A$, negation takes the complete proposition A within its scope. The formulation "it is not the case that A" makes this scope explicit for the reason that A represents a complete proposition, and $\neg$ operates upon that proposition as a whole. By contrast, "is not" occurs grammatically within a particular predicate construction and can make negation appear to be purely a modification of that predicate. Furthermore, "is not" cannot provide a uniform translation for every possible A. If A is "The train has arrived", for instance, $\neg A$ is systematically rendered as "It is not the case that the train has arrived". The same form can be used regardless of the grammatical structure of A. The point is not that ordinary expressions containing "not" are linguistically incorrect, but that "it is not the case that" more transparently represents the formal scope of $\neg$. Thus, whenever $\neg$ occurs in the natural-language examples of this document, it is translated using this formulation.

With this distinction established, the coherent classical relationship between acceptance and rejection can be represented as $+A \leftrightarrow -(\neg A)$ and $-A \leftrightarrow +(\neg A)$.[15] These relations can be displayed implicatively as $+A \vdash -(\neg A)$, $-(\neg A) \vdash +A$, $-A \vdash +(\neg A)$, and $+(\neg A) \vdash -A$ and, within a bilateral natural deduction setting, exhibited as:[16]

[14] This statement provides further grounds for promoting negation to the status of a structural operator and, within sequent calculus, for classifying negation rules alongside structural principles such as reflexivity, weakening, contraction, and cut. We recognise, however, as stated at the outset of this work, that this thesis requires an independent enquiry or, more appropriately, a series of dedicated investigations.

[15] Bear in mind that the identities $+A \leftrightarrow -(\neg A)$ and $-A \leftrightarrow +(\neg A)$ each express two directional inferential relations. In the first identity, $+A \vdash -(\neg A)$, the inference moves from acceptance of A to rejection of $\neg A$, whereas the reverse direction, $-(\neg A) \vdash +A$, inferentially moves from rejection of $\neg A$ back to acceptance of A. In the second identity, $-A \vdash +(\neg A)$, the inference moves from rejection of A to acceptance of $\neg A$, whereas the reverse inferential direction, $+(\neg A) \vdash -A$, moves from acceptance of $\neg A$ back to rejection of A. Thus, each identity is bidirectional since one inference proceeds from one polar position to its negation-mediated counterpart, while the converse inference returns from that counterpart to the original position.

[16] In a future formulation of a unified system of bilateral natural deduction, we prefer to call these rules governing polarity signs "polarisation rules" because their distinctive function is to regulate transitions between the positive and negative poles. The relations $+A \vdash -(\neg A)$, $-(\neg A) \vdash +A$, $-A \vdash +(\neg A)$, and $+(\neg A) \vdash -A$ do not simply introduce or eliminate negation; they systematically coordinate a change in polarity with a corresponding

$$\frac{+A}{-(\neg A)}\,-\neg I \qquad \frac{-(\neg A)}{+A}\,-\neg E \qquad \frac{-A}{+(\neg A)}\,+\neg I \qquad \frac{+(\neg A)}{-A}\,+\neg E$$

These relations connect the two opposite speech-act positions by coordinating a change in polarity with a corresponding negation of propositional content. Thus, the inferential movement between the positive and negative poles does not amount to a direct reversal of polarity toward the same A. Rather, negation mediates that shift, allowing acceptance and rejection to remain formally opposed while standing in a methodical classical relationship.

The complementary notions of coherency and incoherency can now clarify this relationship further without displacing acceptance and rejection from the centre of the analysis. Consider the coherent configurations $+(+A \wedge -(\neg A))$ and $-(-A \wedge +(\neg A))$, which are the bilateral forms of the law of non-contradiction (LNC) and can be derived in a signed sequent calculus setting as follows:[17]

$$\dfrac{\dfrac{+A \Rightarrow +A}{+A, -(\neg A) \Rightarrow}\,-L\neg}{+(+A \wedge -(\neg A)) \Rightarrow}\,+L\wedge \qquad \dfrac{\dfrac{-A \Rightarrow -A}{-A, +(\neg A) \Rightarrow}\,+L\neg}{-(-A \wedge +(\neg A)) \Rightarrow}\,-L\wedge$$

Each derivation originates from a reflexive signed sequent, $+A \Rightarrow +A$ or $-A \Rightarrow -A$. The appropriate signed negation rule then generates the signed-negated counterpart of A, producing respectively $+A$ with $-(\neg A)$ and $-A$ with $+(\neg A)$, before the signed conjunction rule forms the resulting compound.[18] Since the change of polarity is mediated by negation change in propositional content through negation. The term "polarisation rules" therefore captures their specifically bilateral function more precisely than ordinary negation-rule terminology.

[17] A central bilateral note must be made here, and that is that the two configurations $+(+A \wedge -(\neg A))$ and $-(-A \wedge +(\neg A))$, as compound signed formulae, must be read in terms of their external signs. Accordingly, the first is read as the acceptance of the conjunction $A \wedge \neg A$, whereas the second is read as the rejection of the conjunction $A \wedge \neg A$. The internal signs are inactive with respect to the natural-language reading of the compound formula, since $+(+A \wedge -(\neg A))$ should not be read as the acceptance of accepting A and rejecting ¬A, nor should $-(-A \wedge +(\neg A))$ be read as the rejection of rejecting A and accepting ¬A. Nevertheless, these internal signs must be formally preserved because they record the inferential history through which the constituent formulae were derived and thus function as inferential fingerprints of the derivations. In the formal notation of bilateralism (that we are formulating in this manuscript) this principle of preserving the internal signs while reading only the external signs applies to all signed compound formulae, whether they are the displayed bilateral forms of LNC, the bilateral forms of the law of excluded middle (LEM), $-(+A \vee -(\neg A))$ and $+(-A \vee +(\neg A))$, or any other signed compound formulae.

[18] Within the framework of bilateralism naturally proposed by this manuscript, whose complete formulation will be developed in future works, the concept of sign-negation maintains that the polarity signs + and − are methodically linked through the operator of negation. This relationship arises for two complementary reasons. First, acceptance and rejection, represented respectively as positive and negative inferential positions, stand in opposition through the mediating role of negation, since $+A$ is related to $-(\neg A)$, while $-A$ is related to $+(\neg A)$.

from a reflexive starting point, both bilateral forms of LNC preserve reflexivity and constitute coherent configurations. If A represents “The door is locked”, $+(+A \wedge -(\neg A))$ is read as the acceptance of the contradictory conjunction “The door is locked and it is not the case that the door is locked”, whereas $-(-A \wedge +(\neg A))$ is read as the rejection of that contradictory conjunction. The external signs therefore determine the acceptance or rejection of the completed bilateral LNC formulae, while the preserved internal signs record the inferential structure through which their constituents were derived.

Conversely, incoherency can be illustrated through two related kinds of bilateral configurations, namely, the contrary states $+(+A \wedge -A)$ and $-(-A \wedge +A)$, and the pseudo-forms of LNC $+(+A \wedge +(\neg A))$ and $-(-A \wedge -(\neg A))$.[19] Although their defects differ, all four configurations undermine reflexivity and thus disrupt the structured opposition between acceptance and rejection. In the contrary states, the same propositional content A occurs under the opposite polarities $+A$ and $-A$, bringing acceptance and rejection directly to bear upon A without the mediation of negation. Their incoherency is revealed by the signed sequent derivations required to generate them:

$$\dfrac{\dfrac{+\mathbf{A} \Rightarrow +(\neg \mathbf{A})}{+\mathbf{A}, -\mathbf{A} \Rightarrow}\,{-}\mathbf{L}\neg}{+(+\mathbf{A} \wedge -\mathbf{A}) \Rightarrow}\,{+}\mathbf{L}\wedge \qquad \dfrac{\dfrac{-\mathbf{A} \Rightarrow -(\neg \mathbf{A})}{-\mathbf{A}, +\mathbf{A} \Rightarrow}\,{+}\mathbf{L}\neg}{-(-\mathbf{A} \wedge +\mathbf{A}) \Rightarrow}\,{-}\mathbf{L}\wedge$$

The derivations begin respectively with the irreflexive sequents $+A \Rightarrow +(\neg A)$ and $-A \Rightarrow -(\neg A)$, in which A becomes $\neg A$ while its polarity remains unchanged. The subsequent negation rules expose the consequence of this defect by producing $+A,-A \Rightarrow$ and $-A,+A \Rightarrow$, thus placing the contrary polarities directly upon the identical A. The resulting compounds are therefore incoherent because their inferential structures collapse the distinction between the positive and negative positions upon the same propositional content. The pseudo-forms of

---

Second, within the classical bilateral framework, both the signed forms of the rules of double negation introduction (DNI: $+A \Rightarrow +(\neg\neg A)$ and $-A \Rightarrow -(\neg\neg A)$) and double negation elimination (DNE: $+(\neg\neg A) \Rightarrow +A$ and $-(\neg\neg A) \Rightarrow -A$) are valid. Consequently, the interaction between signs and negation is reversible, allowing the positive and negative positions to remain systematically connected through negation while preserving their formal distinction. From these two points, a central corollary concerning the representation of formulae within a bilateral structure can be derived namely, within the formal environment of bilateralism, all exhibited formulae are “signed-negated formulae” (SNFs).

[19] Note that incoherency can equally arise through disjunctive configurations. The contrary disjunctive relations $+(+A \vee -A)$ and $-(-A \vee +A)$ are incoherent because they place acceptance and rejection directly upon the same propositional content without negation mediating their opposition. Their derivations originate respectively from the irreflexive sequents $+(\neg A) \Rightarrow +A$ and $-(\neg A) \Rightarrow -A$, thus violating the reflexive stability required by the bilateral framework. Similarly, the pseudo-forms of LEM, $-(-A \vee -(\neg A))$ and $+(+A \vee +(\neg A))$, are incoherent because A and $\neg A$ occur under the same polarity. Consequently, both contrary disjunctions and pseudo-LEM configurations disrupt the coordinated relation between polarity and negation required for coherent bilateral organisation. However, owing to limitations of space and the relative obviousness of the matter in light of the preceding analysis, we shall not provide their corresponding signed ssequent derivations here.

LNC +(+A ∧ +(¬A)) and −(−A ∧ −(¬A)) exhibit the complementary defect. Although their underlying propositional content has the conjunctive form A ∧ ¬A and thus resembles LNC, their signed organisation prevents them from constituting genuine bilateral forms of LNC. Their attempted sequent derivations are:

$$\dfrac{\dfrac{+A \Rightarrow -A}{+A, +(\neg A) \Rightarrow}\,{+L\neg}}{+(+A \wedge +(\neg A)) \Rightarrow}\,{+L\wedge} \qquad \dfrac{\dfrac{-A \Rightarrow +A}{-A, -(\neg A) \Rightarrow}\,{-L\neg}}{-(-A \wedge -(\neg A)) \Rightarrow}\,{-L\wedge}$$

Here incoherency originates immediately in the irreflexive sequents +A ⇒ −A and −A ⇒ +A, which reverse polarity while leaving the propositional content unchanged. The subsequent rules preserve this initial inferential defect, producing the same-polarity pairs +A with +(¬A) and −A with −(¬A). Thus, whereas the contrary states place opposite polarities upon the same A, the pseudo-forms of LNC place the same polarity upon A and ¬A. These complementary failures distinguish both kinds of configuration from genuine bilateral LNC, whose compound forms derive from the reflexive starting points +A ⇒ +A and −A ⇒ −A and coordinate opposite polarities through negation. As a result, all four configurations are incoherent because their derivational structures fail to preserve the reflexive organisation required for the bilateral opposition between acceptance and rejection.

The bilateral model gives this distinction a precise syntactic form. Within its signed language, propositional content must occur under either positive or negative polarity, so an unsigned A does not by itself constitute a well-formed formula. Since + and − determine polarity while ¬ operates upon propositional content, for every formula, whether elementary or compound, four basic syntactical possibilities arise, namely, +A, −A, +(¬A), and −(¬A). This fourfold structure preserves the distinction −A ≠ ¬A and, more generally, the independence of polarity from negation, since A or ¬A may occur under either polarity. The same syntactic architecture extends to compound signed formulae. As illustrated by +(+A ∧ +B), −(−A ∧ −B), and +(+A ∧ −(¬B)), the external sign determines the polarity and natural-language reading of the completed compound through acceptance or rejection, whereas the internal signs are not separately read as speech acts but preserve the inferential origins of its constituents and become relevant again upon decomposition. This distinction applies generally across logical constants, as shown by the coherent LEM configurations −(+A ∨ −(¬A)) and +(−A ∨ +(¬A)), in which the external sign likewise determines the polarity of the completed disjunction while the internal signs retain its inferential structure. Therefore, the bilateral syntax preserves the distinction between polarity and propositional content at both the elementary and compound levels, while showing that their coherent organisation depends upon their structural coordination with negation rather than upon the particular constant employed.

To conclude, this paper has sought to establish a systematic bilateral account of acceptance and rejection as classical speech acts expressing two formally opposed polar positions toward propositional content. The principal aims stated at the outset have thus been fulfilled by showing that these speech acts can be abstracted from their contingent linguistic circumstances and represented respectively through the positive and negative positions $+A$ and $-A$, while preserving their opposition without identifying rejection with propositional negation. The central identities $+A \leftrightarrow -(\neg A)$ and $-A \leftrightarrow +(\neg A)$, together with their four directional inferential relations $+A \vdash -(\neg A)$, $-(\neg A) \vdash +A$, $-A \vdash +(\neg A)$, and $+(\neg A) \vdash -A$, demonstrate that inferential movement between the two poles is reversible while remaining mediated by negation. Negation is therefore not itself the negative pole represented by rejection; rather, it operates upon propositional content and provides the mechanism through which the two polar positions are methodically related. The supplementary notions of coherency, incoherency, and reflexivity have further clarified the structural conditions governing this relationship. Coherent configurations preserve the opposition between acceptance and rejection by coordinating a change of polarity with negation from reflexive starting points, whereas incoherent configurations disrupt this organisation either by imposing contrary polarities directly upon the same propositional content or by assigning the same polarity to A and $\neg A$. The bilateral forms of LNC and LEM show how this organisation extends respectively across conjunction and disjunction, while the signed derivations of LNC and their contrast with contrary and pseudo-LNC configurations demonstrate how the distinction between coherent and incoherent structures is grounded in their inferential origins. The analysis therefore supports the central distinction between negative polarity and negated propositional content while showing how these formally distinct elements can interact systematically within a classical bilateral setting.

The manuscript has also laid the foundations for the syntax of a formally outlined bilateral system in which every formula, whether elementary or compound, occurs under positive or negative polarity, thus preserving the distinction $-A \neq \neg A$. The four syntactical possibilities $+A$, $-A$, $+(\neg A)$, and $-(\neg A)$ express the independence of polarity from negation, while their extension to compound signed formulae establishes a corresponding distinction between external and internal signs. The external sign determines the polarity and natural-language interpretation of the completed compound through acceptance or rejection, whereas the internal signs are not independently read as speech acts but must nevertheless be preserved as inferential fingerprints recording the origins of its constituents and becoming relevant again upon decomposition. These results provide a foundation for a number of directions of future research already suggested by the present analysis. The bilateral system of acceptance and rejection developed here may be employed in formulating bilateral frameworks for a theory of logical consequence and a theory of meaning. The mediating function of negation also provides grounds for a separate investigation into whether negation should be understood not merely as a logical constant alongside conjunction, disjunction, and implication, but as a structural operator responsible for regulating polarity and thus contributing to the dual character of classical logic. A further area of research concerns the construction of a specifically bilateral account of warrant. Since acceptance and rejection constitute opposed inferential positions and coherency determines how inferential movement between them may be appropriately mediated by negation, these polar notions may provide the basis for developing a bilateral conception of inferential warrant in which entitlement is articulated through relations between positive and negative positions rather than through either position alone. Each of these possibilities requires independent and substantially more extensive investigation. The present paper is therefore intended to provide a foundational and boundary-setting study for subsequent research and, ultimately, for the development of a more comprehensive formulation of bilateralism grounded in the polar architecture of acceptance and rejection.